\documentclass[11pt,twoside]{article}

\usepackage{asp2004}
\usepackage{epsf}
\usepackage{psfig}
\usepackage{lscape}

\begin{document}
\title{Turning AGN Microlensing From a Curiosity Into a Tool }   
\author{C.S. Kochanek, X.Dai, C. Morgan, N. Morgan, S. Poindexter}   
\affil{Department of Astronomy, The Ohio State University}    
\author{G. Chartas}
\affil{Department of Astronomy and Astrophysics, Pennsylvania State University}    

\begin{abstract} 
Microlensing of gravitationally lensed quasars by the stars in the foreground
lens galaxy can be used to probe the nature of dark matter, to determine the mean
stellar mass in the lens galaxy, and to measure the internal structure of quasar
accretion disks.  Until recently, little progress has been made toward using
microlensing for these purposes because of the difficulty in obtaining the necessary
data and the lack of good analysis methods.  In the last few years, both problems
have been solved.  In particular, Bayesian analysis methods provide a general
approach to measuring quantities of physical interest and their uncertainties from 
microlensing light curves.  We discuss the data and the analysis methods and show
preliminary results for all three astrophysical applications.
\end{abstract}


\section{Introduction}   

Suppose I advertise to you a set of observations that simultaneously 
probe the nature of dark matter, the mean stellar mass in cosmologically distant
galaxies, and resolves the internal structure of quasars.  This sounds like
a marvelous scientific opportunity.  Yet despite having the capability of carrying
out these observations for over a quarter of a century, there are few quantitative
results.

The observations are the monitoring of gravitationally lensed quasars for the
effects of microlensing by the stars in the lens galaxy.  In a gravitational
lens, we see (usually) two or four images of a background quasar created by 
the deflection of light in the gravitational field of a foreground galaxy. 
If we then monitor the brightness of the quasar images, we will observe them
to vary due to two effects.  First, if the luminosity of the quasar varies,
we will see those variations in the individual images but with temporal
shifts due to the different light travel times associated with each image.
These time delays can be measured by cross correlating the light curves of the 
individual images, and they measure the surface mass density of the lens 
galaxy at the radius of the images from the center of the lens galaxy.
Second, we will see uncorrelated variations in the brightness of the images
due to {\it microlensing} by the stars in the lens galaxy near each image
of the quasar.   Our advertisement was for the scientific potential
of microlensing.

Two basic problems have blocked delivering on the advertisement.  The first problem
is sociological.  Time on optical telescopes is generally 
allocated by the night, and this sociological choice makes it nearly impossible 
to monitor gravitational
lenses.  A scientific program to study variability in an interesting sample
of lenses requires roughly 1~hour/night, {\it every night}, on a 1--2m class 
telescope.   Since the time-allocation scheme does not allow for such
studies,  monitoring lenses has been restricted to small numbers of lenses
and short periods of time.
The second problem is statistical.  While microlensing variability 
depends on the nature of dark matter, the mean stellar mass in the lens, and the internal
structure of quasars, it is not amenable to simple
analyses.   Even for the few lenses that were successfully
monitored and showed microlensing variability, there was no good way to
analyze the results without making compromising assumptions.

In this brief review, we outline the physics of microlensing in \S2, the nature of
the data in \S3, and the Bayesian Monte Carlo method we have
developed to analyze the data in \S4.  In \S5 we present results on the
size of quasars, the fraction of mass in stars and the mean mass of the stars.
In \S6 we outline some of the physical,
computational and statistical issues of our approach and how 
they might be addressed.  In \S7 we 
discuss the future of the method.  Wambsganss (2006) provides a complete
review of the microlensing field and Kochanek (2006) provides a review of
strong lensing.  We use a flat $\Omega_0=0.3$ cosmology and a 
a Hubble constant of $H_0 = 100 h$~km~s$^{-1}$~Mpc$^{-1}$.

\begin{figure}[t]
\centerline{
  \psfig{figure=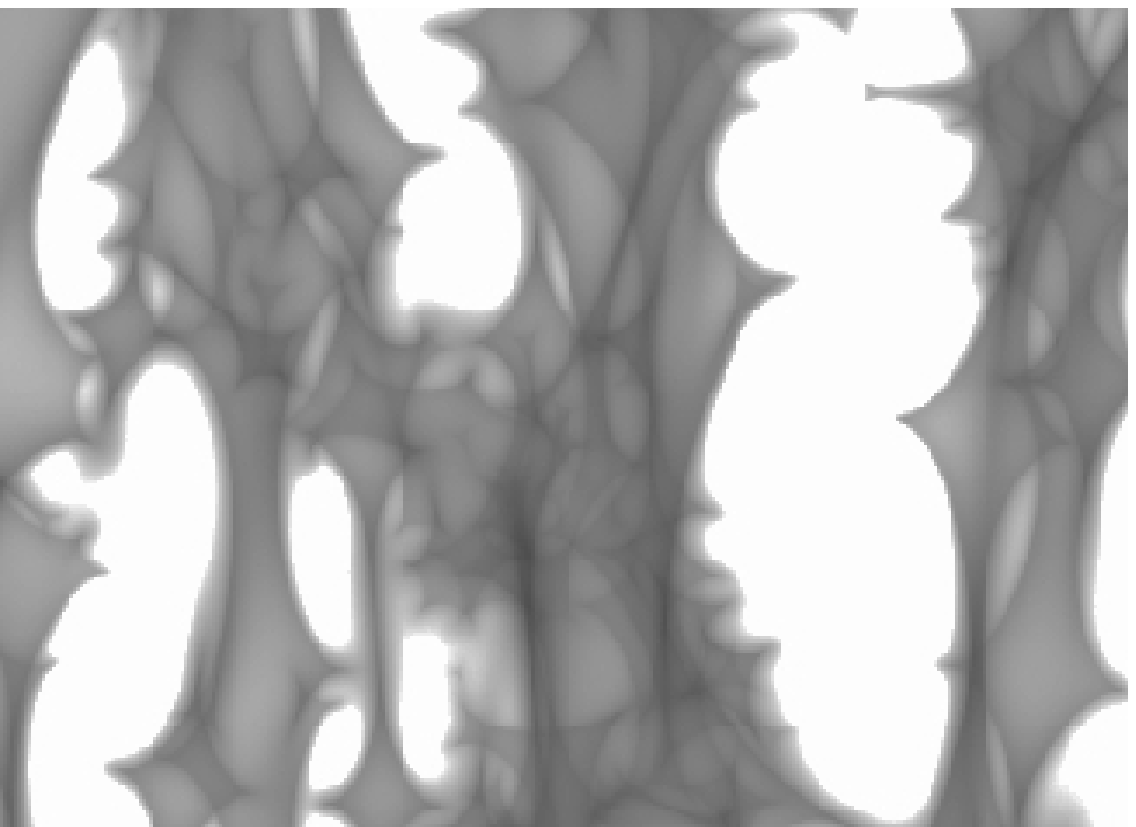,width=2.5in}
  \psfig{figure=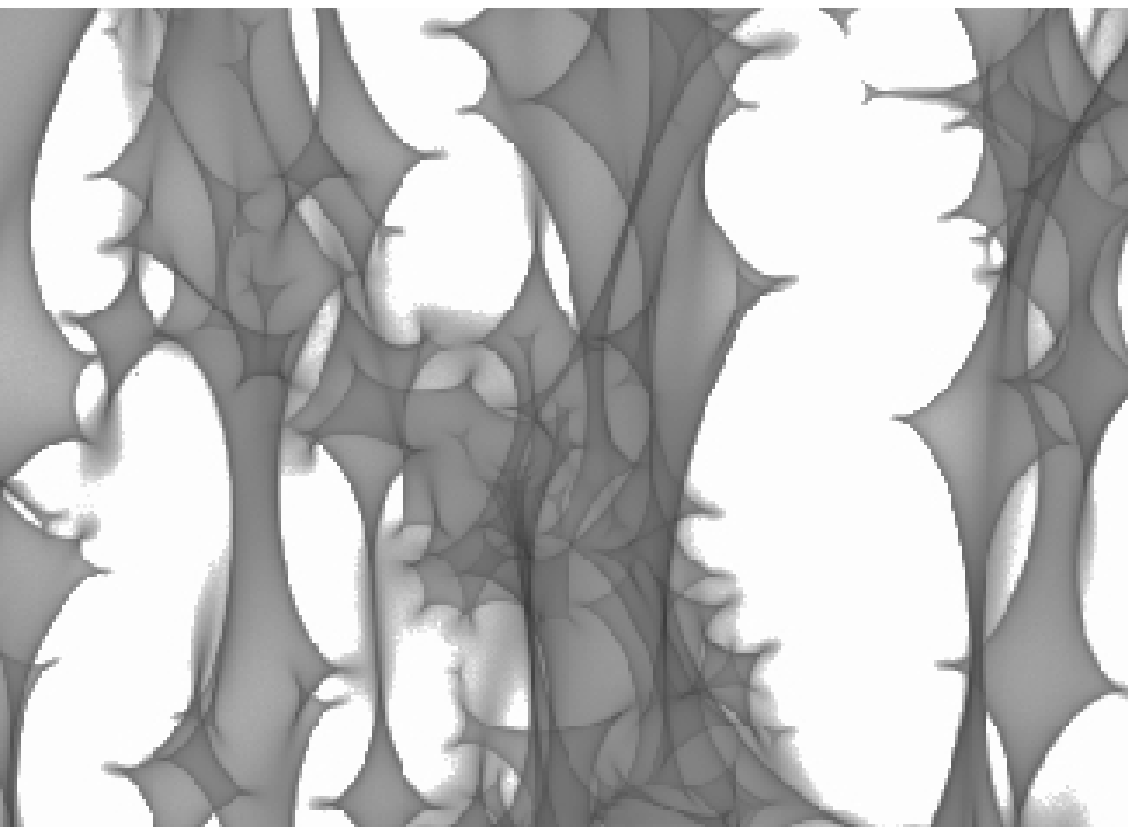,width=2.5in}
   }
\caption{ A typical magnification pattern for image A of Q2237+0305.  The left panel is smoothed
  to the source size needed to fit the optical light curves.  The right panel shows
  the same pattern for a source three times smaller.  Darker regions correspond to
  higher magnifications.  The peak magnifications drop from a factor of 10 (right panel)
  to a factor of 6 (left panel).  The center of the lens galaxy lies to the right.
  } 
  \label{fig:map}
\end{figure}

\section{The Basics Of Microlensing}

The region near each lensed quasar image is characterized by three dimensionless numbers:
the mean surface density $\kappa$, the surface density in stars $\kappa_*\leq\kappa$
and the (tidal) shear $\gamma$.  The surface densities are in units of the critical
surface density for gravitational lensing.  The surface density and the shear
determine the two inverse magnification eigenvectors $1-\kappa \pm \gamma$, whose
values determine many of the large scale statistical properties of microlensing.
The stars, or microlenses,
locally deflect light rays and this leads to a complicated magnification pattern   
whose characteristic scale is set by the Einstein radii of the stars
projected onto the source plane
\begin{equation}
     R_E = D_{OS} \left[ { 4 G \langle M \rangle \over D_{OL} c^2 } { D_{LS} \over D_{OS} } \right]^{1/2}
         =  \left( 5.5 \times 10^{16} \right) \left[ { \langle M \rangle \over h M_\odot } \right]^{1/2} \hbox{cm}
     \label{eqn:rein}
\end{equation}
where $\langle M\rangle$ is the mean mass of the stars and we used the angular diameter 
distances $D_{OL}$, $D_{OS}$ and $D_{LS}$ between the observer, lens and source for the lens PG1115+080.
Fig.~\ref{fig:map} shows some typical magnification patterns for the lens Q2237+0305.
While $R_E$ sets a scale for the features, the surface density
of stars ($\kappa_*$), their relative positions, and the local magnification ($\kappa$,
$\gamma$) all interact to create complex patterns with networks of caustic curves
on which the magnification is formally divergent (see Fig.~\ref{fig:map}).      

These patterns are the magnifications experienced by a point source.  We observe
the fluctuations found by convolving the pattern with the surface brightness of
the source -- larger sources will show steadily smaller fluctuations.  In Fig.~\ref{fig:map}
we have actually shown models convolved with a source with the typical size
needed to fit the optical microlensing variability and with a source three times
smaller.
Quasars
are extended sources with a minimum size scale of order the gravitational radius
of the black holes powering quasars
\begin{equation}
    r_g =  { G M_{BH} \over c^2 } = \left(1.5 \times 10^{14}\right) \left( { M_{BH} \over 10^9 M_\odot} \right) \hbox{cm}.
     \label{eqn:rgrav}
\end{equation}
For a standard thin accretion disk model of a quasar (Shakura \& Sunyaev 1973), the black hole is surrounded
by a thermally radiating disk with a temperature profile of
\begin{equation}
     T(R)^4 =  { 3 G M_{BH} \dot{M} \over 8 \pi R^3 \sigma } \left[ 1 - \left( {R_{in} \over R} \right)^{1/2} \right] 
     \label{eqn:tdisk}
\end{equation}
where $\dot{M}$ is the mass accretion rate and $R_{in}$ is the inner edge of the accretion 
disk.  The inner edge is in the range of $r_g < R_{in} < 6 r_g$ set by the last
stable orbits for maximally rotating and non-rotating black holes.
Observations of the disk at rest-frame wavelength $\lambda$ correspond to a typical
temperature $k T_\lambda = h c/\lambda$, so we can rewrite the leading part of the temperature
scaling as $T= T_\lambda(R_\lambda/R)^{3/4}$ where
\begin{equation}
      R_\lambda = { 1 \over \pi^2 } \left( { 45 \over 16} { \lambda^4 r_g \dot{M} \over  h_p} \right)^{1/3}
           = (9.7 \times 10^{15}) \left( { \lambda \over \mu m} \right)^{4/3} \left( { M_{BH} \over 10^9 M_\odot }\right)^{2/3}
                \left( { L \over \eta L_E } \right)^{1/3} \hbox{cm}
     \label{eqn:rdisk}
\end{equation}  
and we have replaced the mass accretion rate $\dot{M}$ by the luminosity $L$ relative to the Eddington 
luminosity $L_E$ combined with the radiative efficiency $\eta$ of the accretion, $L = \eta\dot{M}c^2$. 
When observed face-on, the disk has a surface brightness profile of 
\begin{equation}
     f_\nu(R) = { 2 h_p c \over \lambda^3 } \left[ \exp\left({ h_p c \over k T(R) \lambda} \right) - 1 \right]^{-1} 
              = { 2 h_p c \over \lambda^3 } \left[ \exp\left({ R \over R_\lambda} \right)^{3/4} - 1 \right]^{-1}, 
     \label{eqn:fnu}
\end{equation}
where in the latter expression we neglect the central temperature depression of Eqn.~(3) that is
created by the inner edge of the disk.  Numerical experiments
indicate that this approximation is safe for the moment.  Thus, the scale
$R_\lambda \propto \lambda^{4/3}$ in Eqn.~(4) is the characteristic size of the disk at wavelength $\lambda$
because it is the point where the surface brightness of the disk begins to drop exponentially 
as the disk temperature drops below the photon energy.  We  
may also need to worry about viewing angles, the degree to which the model
breaks down very near the black hole as relativity becomes more important,
and non-thermal emission (X-rays or radio).  

We see fluctuations in the brightness because nothing is static.  We are moving, the lens 
galaxy is moving, the stars in the lens galaxy are moving and the quasar is moving.  Since
the patterns associated with each image are different, we see microlensing as uncorrelated
variations in the image fluxes.  Fig.~2 shows an example of a light curve. 
Most studies have held the patterns fixed,
ignoring the motions of the stars, and generated the fluctuations expected from a
linear track across the fixed patterns.  The consequences of the motions of stars
have been considered theoretically (Kundic \& Wambsganss 1993, Schramm et al. 1993,
Wyithe et al. 2000a), and we will return to this point in \S6.
If we ignore the stellar motions, the bulk motions of the observer, lens and
source can be reduced to a net effective velocity $v_e$ across the magnification
pattern. 

One important point to understand is that observations of the lens can only measure
things in ``Einstein'' units, where the length scale is proportional to $R_E \propto \langle M\rangle^{1/2}$
(Eqn~\ref{eqn:rein}).  We can convert to 
physical units only if we have a prior on some other quantity like the 
effective velocity, the mean stellar mass or the size of the quasar.

We are interested in the ``physical'' variables associated with the structure
of the lens ($\kappa$, $\kappa_*$, $\gamma$), the mass of the stars ($\langle M\rangle$),
the structure of the quasar ($R_\lambda$) and the velocities ($v_e$).  
We are not interested in nuisance variables such as the exact distribution
of the stars or the position of the quasar on the pattern.  The challenge 
is to analyze a set of light curves with arbitrary sampling at an assortment of
wavelengths to obtain estimates of physically interesting variables. 

\section{Obtaining the Data}

The major problem in collecting the data is sociological.  The
physical requirement is a 1-2m class telescope that routinely produces images with
1\farcs0 resolution.  Data must be obtained almost every night in order to 
monitor large numbers of lenses varying over a broad range of time scales.
There are few lenses for which we can tolerate gaps of more than 1--2 weeks during 
an observing season.  The
intrinsic variability of the quasar is generally on shorter time scales than the
microlensing variability, but to remove the intrinsic variability you need to 
measure the time delays and correct for them.  Our
present program, covering about 20 lenses during the course of the year with good
coverage in the R-band and sparser coverage in the J, I and B-bands,
requires approximately 1 hour per night.  This is largely
obtained with the queue scheduled SMARTS 1.3m telescope at CTIO, with
considerably less data from Northern sites like APO, FLWO and MDM.

Once you have a well-sampled light curve at one wavelength, you can use more sparsely
sampled light curves at other wavelengths in order to measure the wavelength-dependence
of the variability.  The ability  to use sparsely sampled data at other bands is particularly
important when extending the coverage to wavelengths that can only be studied with
spacecraft -- HST is needed to study the true ultraviolet, and Chandra is needed to
study X-rays.  These wavelengths are particularly interesting because they should
probe regions closer to the inner edge of the accretion disk than the ground-based optical data.

\section{The Bayesian Monte Carlo Method}

Given data $D$, the measured light curves, and model parameters $p$ we can produce a model
light curve and compute how well it fits the observed light curve with a $\chi^2$ 
statistic to get the probability of the data given the parameters, $P(D|p)=\exp(-\chi^2/2)$.  Using Bayes
theorem and estimates of the prior probabilities $P(p)$ for the parameters,
we can estimate the probability distributions for the parameters given the data, 
$P(p|D) \propto P(D|p)P(p)$, with the usual normalizations to make
the total probability equal to unity.  This applies to any microlensing
data we acquire.

Our basic approach is to generate random magnification patterns for a 
range of physical parameters ($\kappa$, $\kappa_*$, $\gamma$, $\langle M\rangle$),
convolve them with a range of randomly selected disk models ($R_\lambda$), randomly
generate light curves (the effective source velocity $v_e$ and nuisance variables 
such as the starting point of the light curve in the pattern), compute the
$\chi^2$ values for each trial light curve and then use Bayes' theorem 
to combine all the results for the individual trials in order to
determine the probability distributions for the physical parameters --
essentially we do the integrals needed for the Bayesian analysis as a large 
Monte Carlo integral.   The only surprise about the approach is that
such a brute force method can successfully generate significant numbers
of model light curves that are statistically acceptable fits to the 
observed light curves. 

Everything we can measure from a microlensing light curve will be in
``Einstein'' units where lengths scale as $\langle M \rangle^{1/2}$.
In order to convert from Einstein units to physical units, we require
a prior on some variable with dimensions.  Our three possible choices
are the mean mass itself, $\langle M\rangle$, the effective velocity $v_e$, 
and the accretion disk size, $R_\lambda$.  Each of these variables
has a reasonable physical prior. We know the mean stellar mass in
our galaxy is (generously) in the range 
$0.1M_\odot < \langle M \rangle < 1.0 M_\odot$, and this is a 
good choice for the prior because the uncertainty in the length scale
depends only on the square root of the mass scale.  If we assume an
accretion disk model, then we can define a prior for the size of
the accretion disk because the observed flux of the quasar 
determines the disk size if we understand the structure and
emission mechanism of the disk.  For example, in a standard thin disk
with no inner edge,   
\begin{equation}
 R_\lambda \simeq 
   { 1.2 \times 10^{15} \over \sqrt{\cos i}} \left( { H_0 D_{OS} \over c} \right) 
    \left( {\lambda_{obs} \over \mu\hbox{m}} \right)^{3/2} 10^{-0.2(I-19)} h^{-1} \hbox{cm}
\end{equation}
where $I$ is the (magnification corrected) I-band magnitude of the quasar and $i$ is the 
disk inclination angle.
In many ways, however, $\langle M \rangle$ and $R_\lambda$
are the most interesting physical variables in the problem,
so it would be better to assume very broad (or no) priors for 
these variables.  

The last possibility is the effective velocity,
which depends on our motion, the peculiar velocity of the lens
and quasar host galaxies, and the local velocities of the stars
in the lens galaxy.  We know our motion well because it is simply
the projection of the CMB dipole velocity onto the source plane 
of the lens.  For most lens galaxies, we know the local velocity
dispersion of the stars reasonably well either from the image
separations in the lens or by direct measurement.  Ideally we
include this directly by allowing the stars to move, but it 
can be mimicked statistically (Wyithe et al. 2000a).  The biggest
uncertainties are the unknown transverse peculiar velocities of the
lens and quasar host galaxies, with the former being
the most important because peculiar velocities grow with time.
Here we must use a statistical model based on N-body simulations
of structure formation, possibly modified when we know that the
lens is in a group or cluster and should have a higher than 
random peculiar velocity.   

\begin{figure}[t]
\centerline{
  \psfig{figure=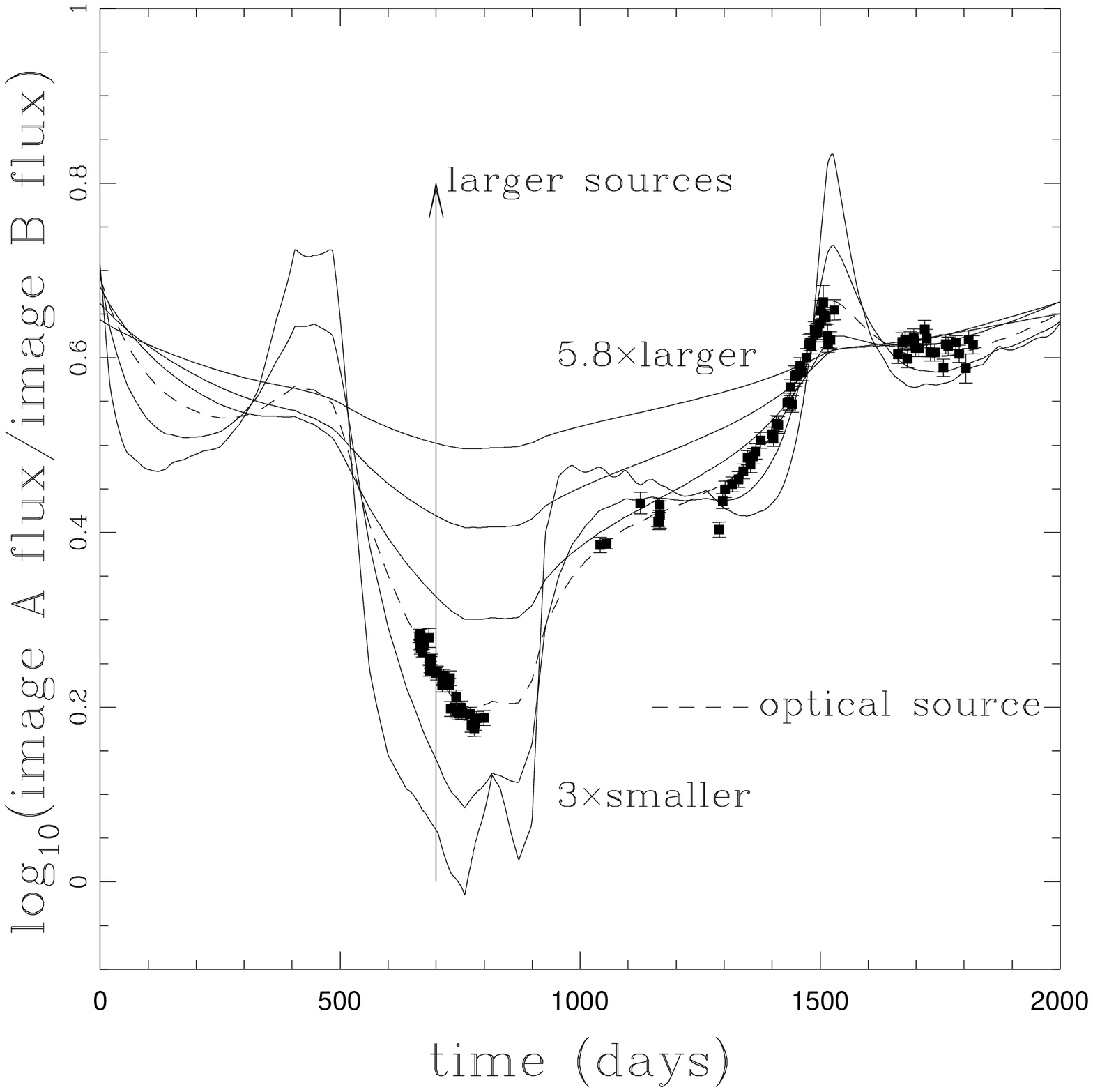,width=2.5in}
  \psfig{figure=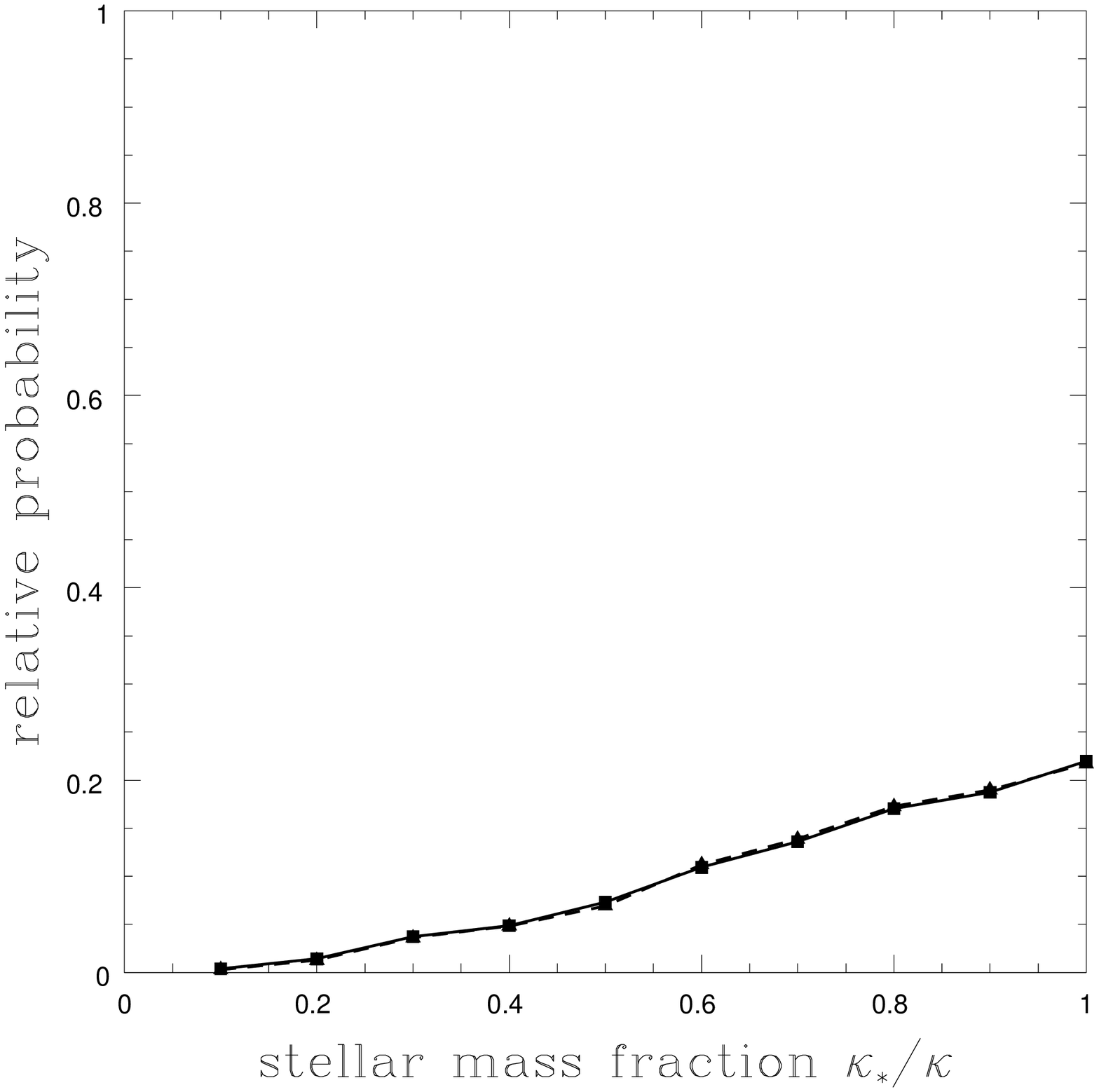,width=2.5in}
  }
\caption{ Model light curves for Q2237+0305.  The dashed curve is a light curve generated
by the Bayesian analysis that fit the OGLE V-band data (points) well and is then extrapolated 
across the magnification pattern.  The solid lines show the effects of changing the source size 
by 0.25 dex (a factor of 1.7) per curve.  Larger sources
more heavily smooth the magnification pattern and show less variability, and vice versa for
small sources.  For a $T\propto R^{-3/4}$ disk, the K-band and 2500\AA\ sources should be
roughly 5 times larger/smaller than the optical source (i.e. three steps from the dashed 
curve).}
\label{fig:lcurve}
\caption{ The fraction of the surface density of Q2237+0305 in stars, $\kappa_*/\kappa$, based
   on the full OGLE light curves.  The likelihood for $\kappa_*/\kappa \sim 1$ has increased
   markedly compared to our original result in Kochanek (2004) with the addition of the 
   OGLE-III light curve data.   Unlike the typical gravitational lens, where we expect
   $\kappa_*/\kappa \sim 0.1$--$0.2$, Q2237+0305 should have $\kappa_*/\kappa \sim 1$ 
   because we see the quasar images through the bulge of a low redshift spiral galaxy.
   }
  \label{fig:kstar}
\end{figure}

\begin{figure}[t]
\centerline{
  \psfig{figure=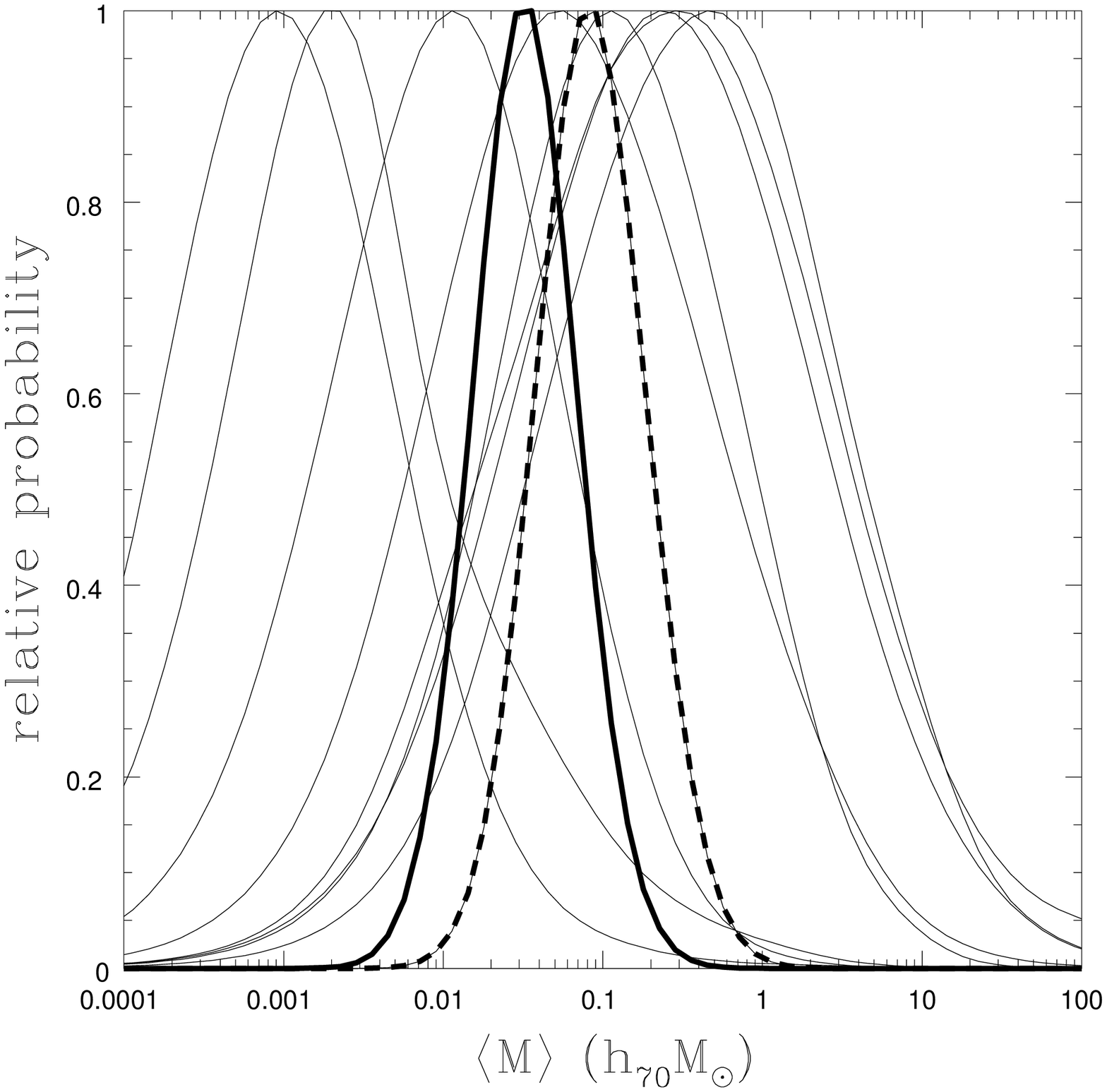,width=2.5in}
  \psfig{figure=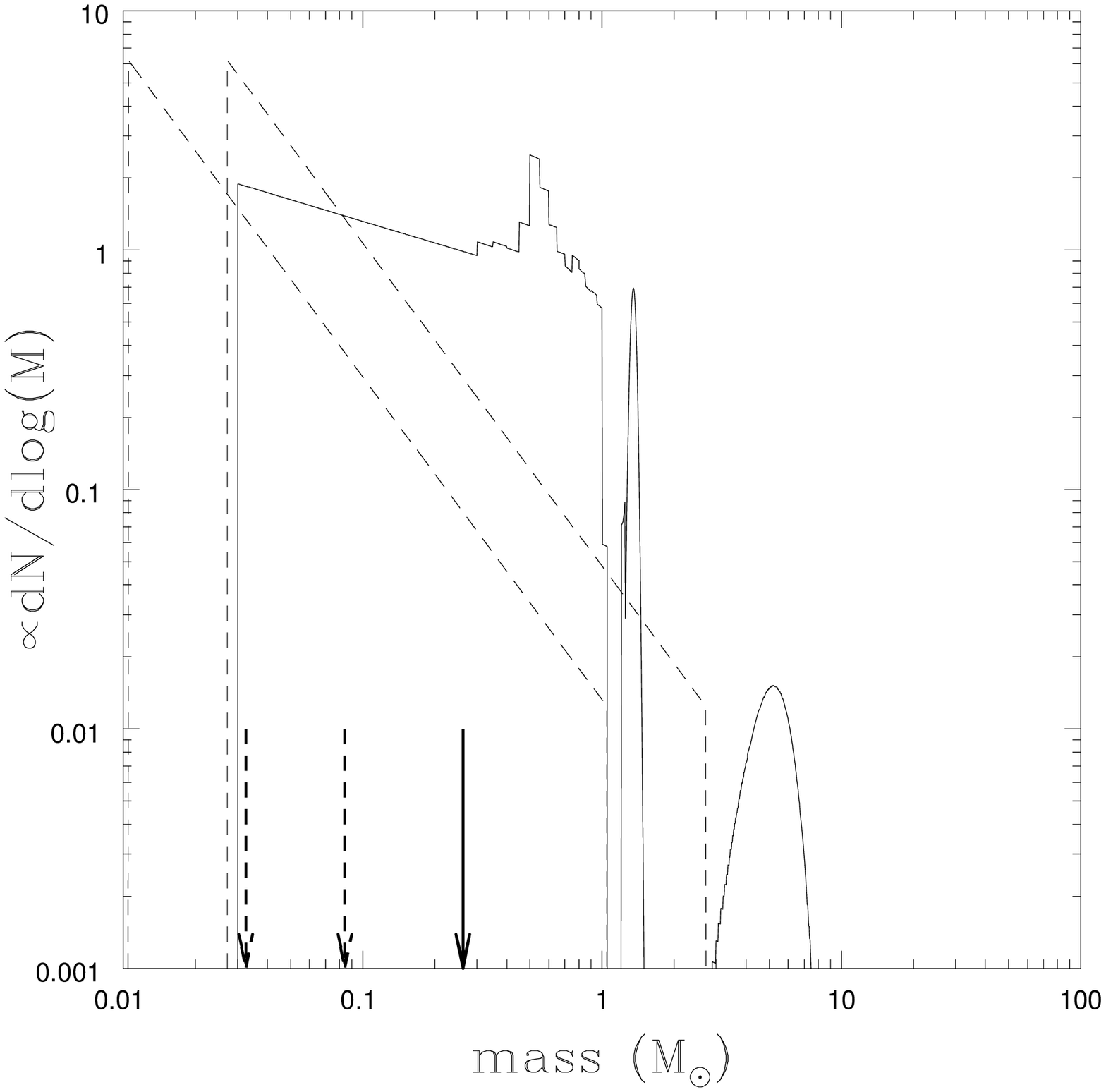,width=2.5in}
  }
\caption{ The mean mass $\langle M \rangle$ of the microlenses for 8 systems (light curves) and
  the joint probability (heavy solid curve).  The heavy dashed curve shows the effect of dropping
  the two systems favoring the lowest masses.  Dropping the two systems favoring the highest
  masses leads to a similar shift, but towards lower masses.   The uncertainties in the results
  for the individual systems are roughly equally due to the model (prior) for the physical 
  velocities and the uncertainties in the effective velocities from fitting the data. 
  }
\label{fig:mass}
\caption{ A comparison of our model mass functions (dashed lines) to a realistic mass function 
   for the Galactic disk (solid curve) by Gould (2000).  The arrows show the mean masses for
   the three mass functions, and the two microlensing mass functions correspond to the two
   cases shown in Fig.~\ref{fig:mass}.  The three peaks in the Gould (2000) mass function correspond
   to white dwarfs, neutron stars and black holes.  Theoretical studies of microlensing strongly
   suggest that our estimates are not sensitive to the shape of the mass function, but this
   comparison suggests that we should shift to a flatter mass function than the Salpeter
   form typically used in microlensing calculations to allow for cleaner comparisons.
   }
  \label{fig:mfunc}
\end{figure}

\begin{figure}[t]
\centerline{
  \psfig{figure=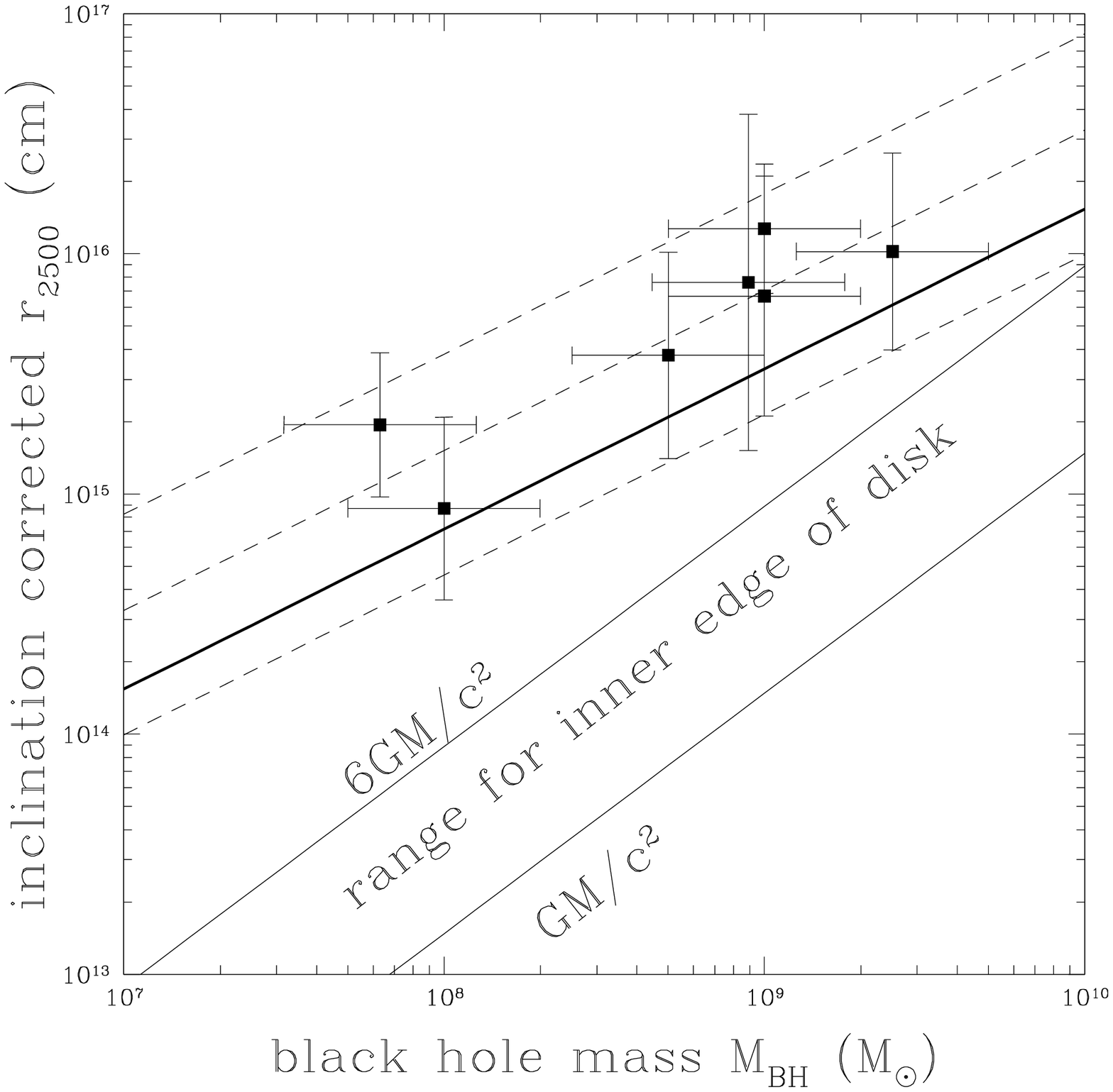,width=2.5in}
  \psfig{figure=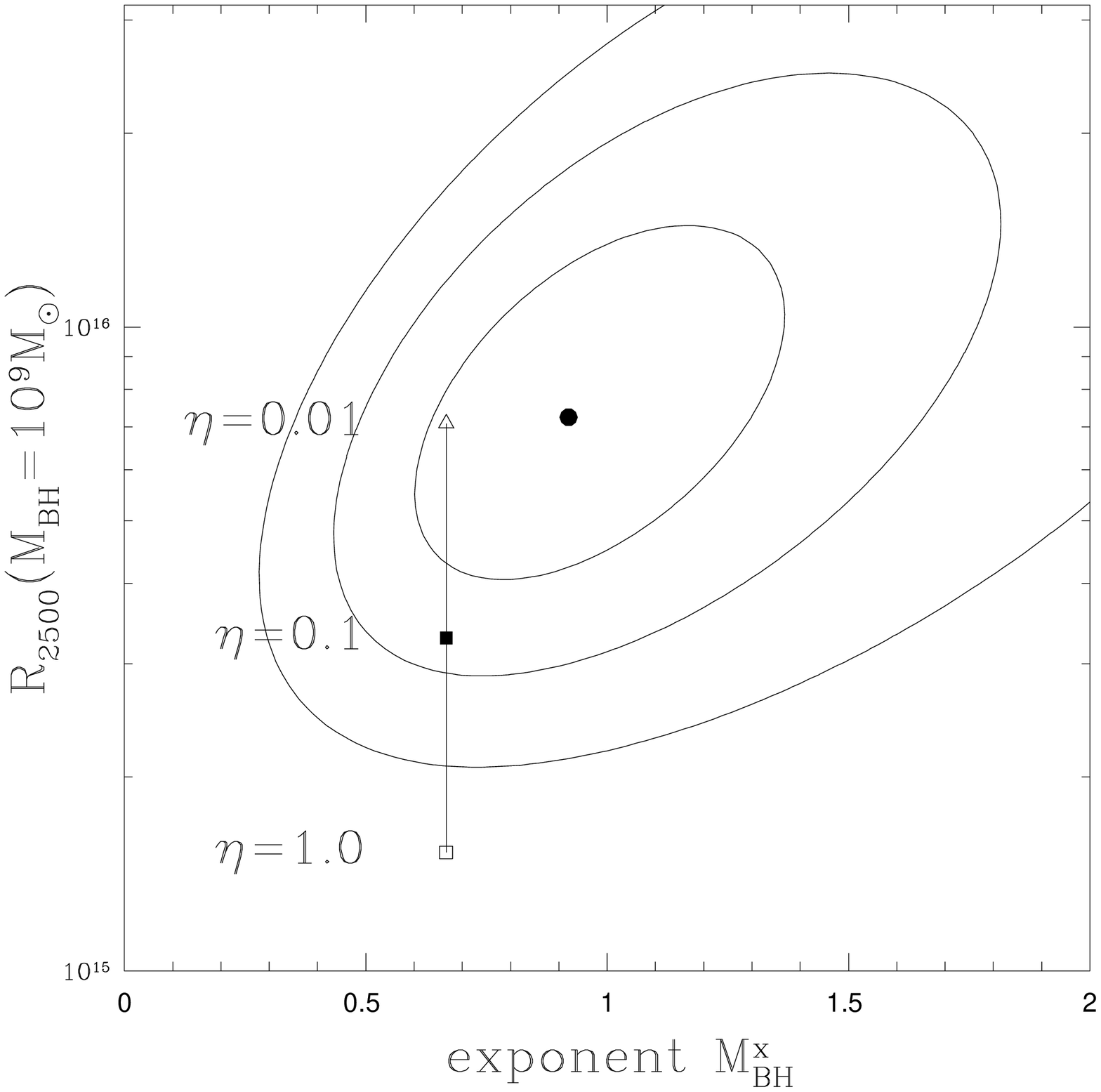,width=2.5in}
  }
\caption{(Left) Accretion disk size versus black hole mass at 2500\AA\ in the rest
  frame of the quasar assuming the mean disk inclination.  This roughly corresponds
  to our standard R-band monitoring filter.  The expectation
  for a standard thin disk and a quasar radiating at the Eddington limit
  with an efficiency of $\eta=0.1$
  is shown by the heavy solid line.  The dashed lines show the expected
  sizes for a $z=1.5$ quasar at the K (top), H and B-bands to illustrate
  the expected dependence on wavelength.  Remember that size ratios can be more 
  accurately measured than absolute sizes.  The lines at $r_g=GM_{BH}/c^2$ 
  to $6 r_g $ show the expected range for the inner edge of the accretion
  disk.  }
  \label{fig:dsize}
\caption{ (Right) A power-law fit to the accretion disk size as a function of black
  hole mass in Fig.~\ref{fig:dsize}.  The contours show the 1-3$\sigma$ limits on
  the two variables.  The points connected by the vertical line show the expectation
  for thin disk models radiating at Eddington with efficiencies of $\eta=0.01$, $0.1$
  and $1.0$.  The estimates from the microlensing data are mildly consistent with
  the theoretical model unless the radiative efficiency is very low.} 
\label{fig:diskfit}
\end{figure}

\begin{figure}[p]
\centerline{
  \psfig{figure=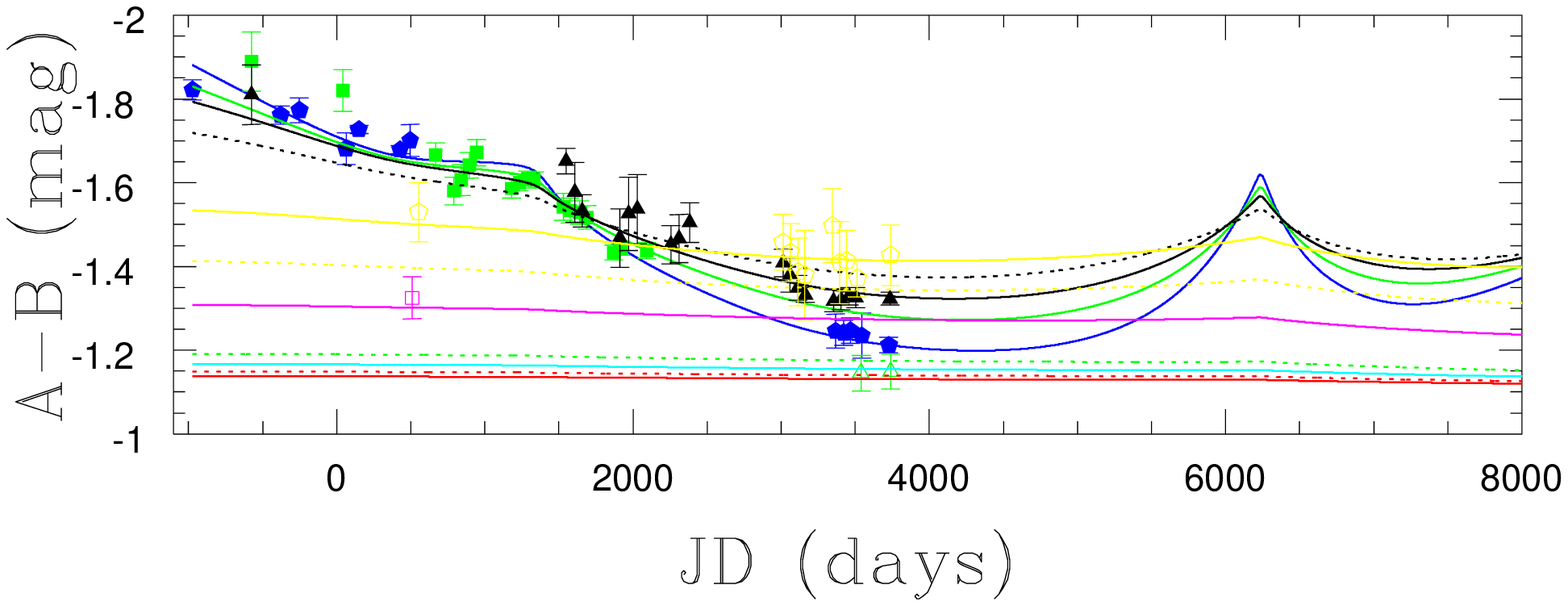,width=5.0in}
  }
\caption{ The HE1104--1805 light curves.  The curves show a model for the BVRIJHK and IRAC Channel 1 (3.6$\mu$m)
  to 4 (8$\mu$m) flux ratios with B-band at the top and 8$\mu$m on the bottom at the beginning of the
  time series.  The points are 90-day running averages of the B (pentagon), V (square), R (triangle), 
  J (open pentagon), K (open square) and 3.6$\mu$m (open triangle) flux ratio measurements.  The
  I-band, H-band, the other IRAC channel data are not shown (but are consistent with the model), and
  only points on the running average separated by at least 45 days are plotted.  Note the optical
  color reversal between the early periods and the present epoch.  The uncertainties in the 
  disk structure estimate shown in Fig.~\ref{fig:he1104like} are driven by the lack of near-IR data 
  the two anomalous V-band data points in the early phases.  The history includes data from
  Remy et al. (1998), Gil-Merino et al. (2002), Ofek \& Maoz (1993) Schechter et al. (2003)
  and Wyrzykowski et al. (2003) as well as our own data.
  }
\label{fig:he1104lc}
\centerline{
  \psfig{figure=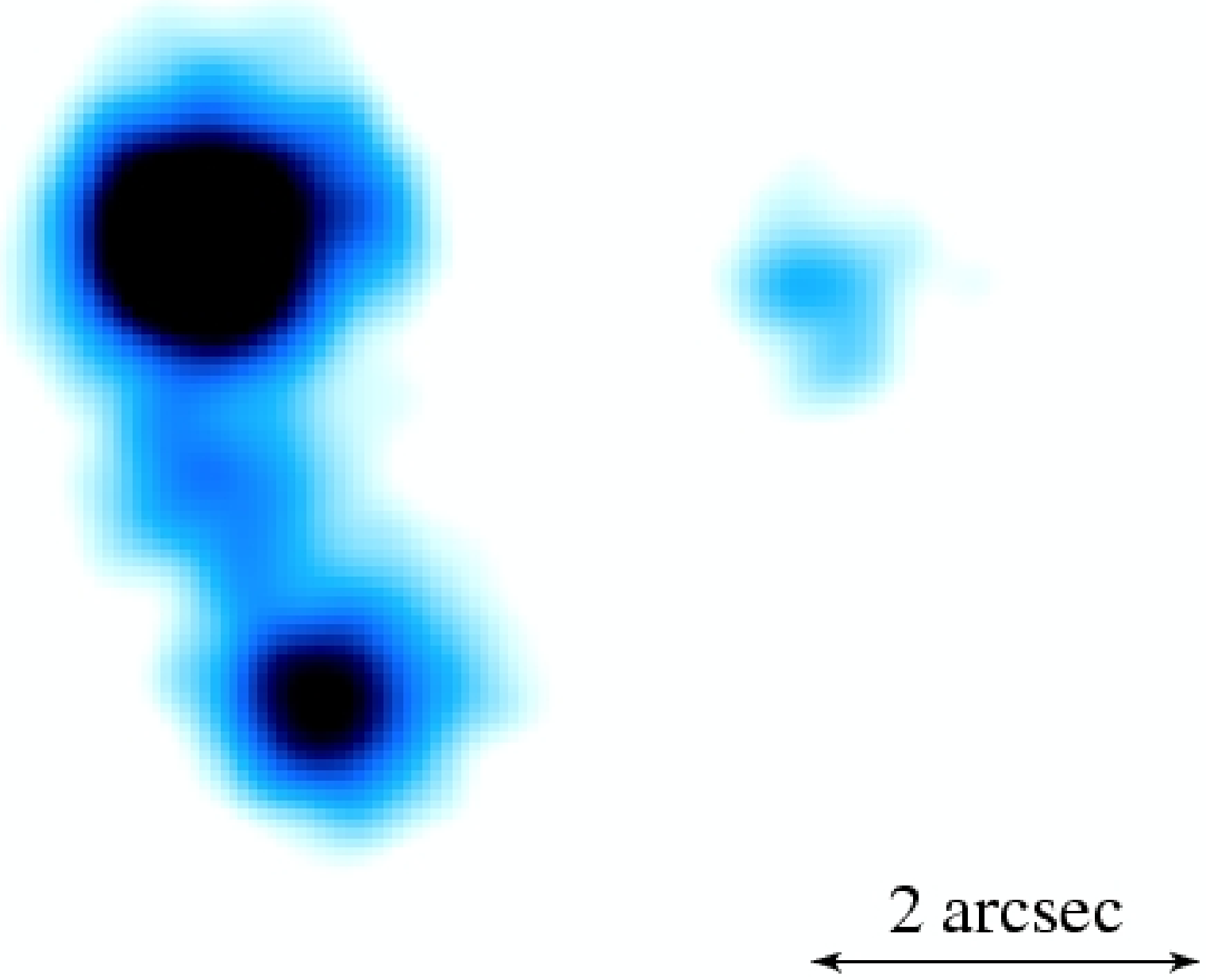,width=2.5in}
  \psfig{figure=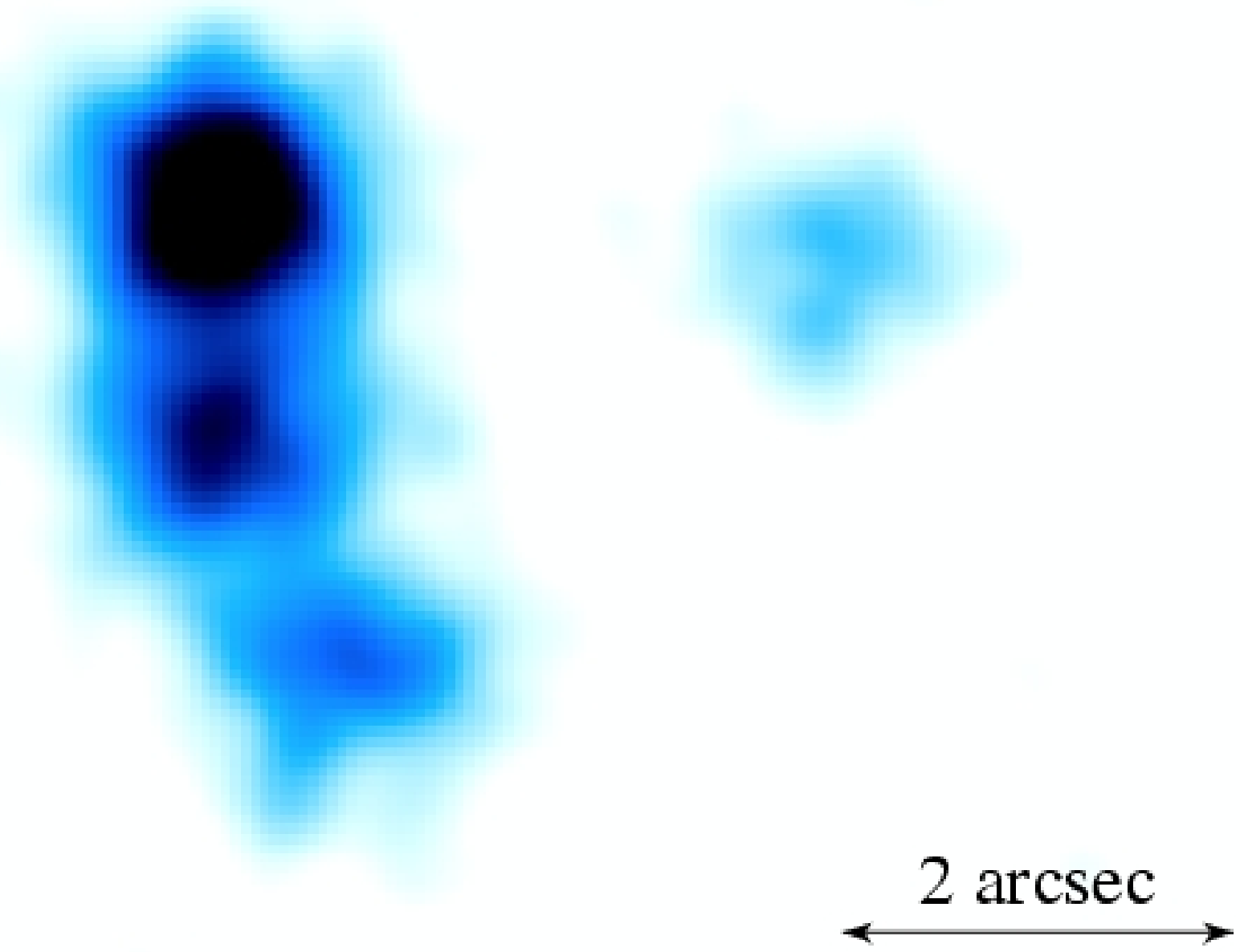,width=2.5in}
  }
\caption{ Two epochs of X-ray data for RXJ1131--1231.  The left panel is the Blackburne et al. (2006) data
  from 12 April 2004, and the right panel is our 10 March 2006 image.  Notice the dramatic
  changes in the A-C X-ray flux ratios produced by microlensing.  The optical flux ratios are
  very different for both epochs.  
  }
\label{fig:ximage}
\end{figure}

\begin{figure}[t]
\centerline{
  \psfig{figure=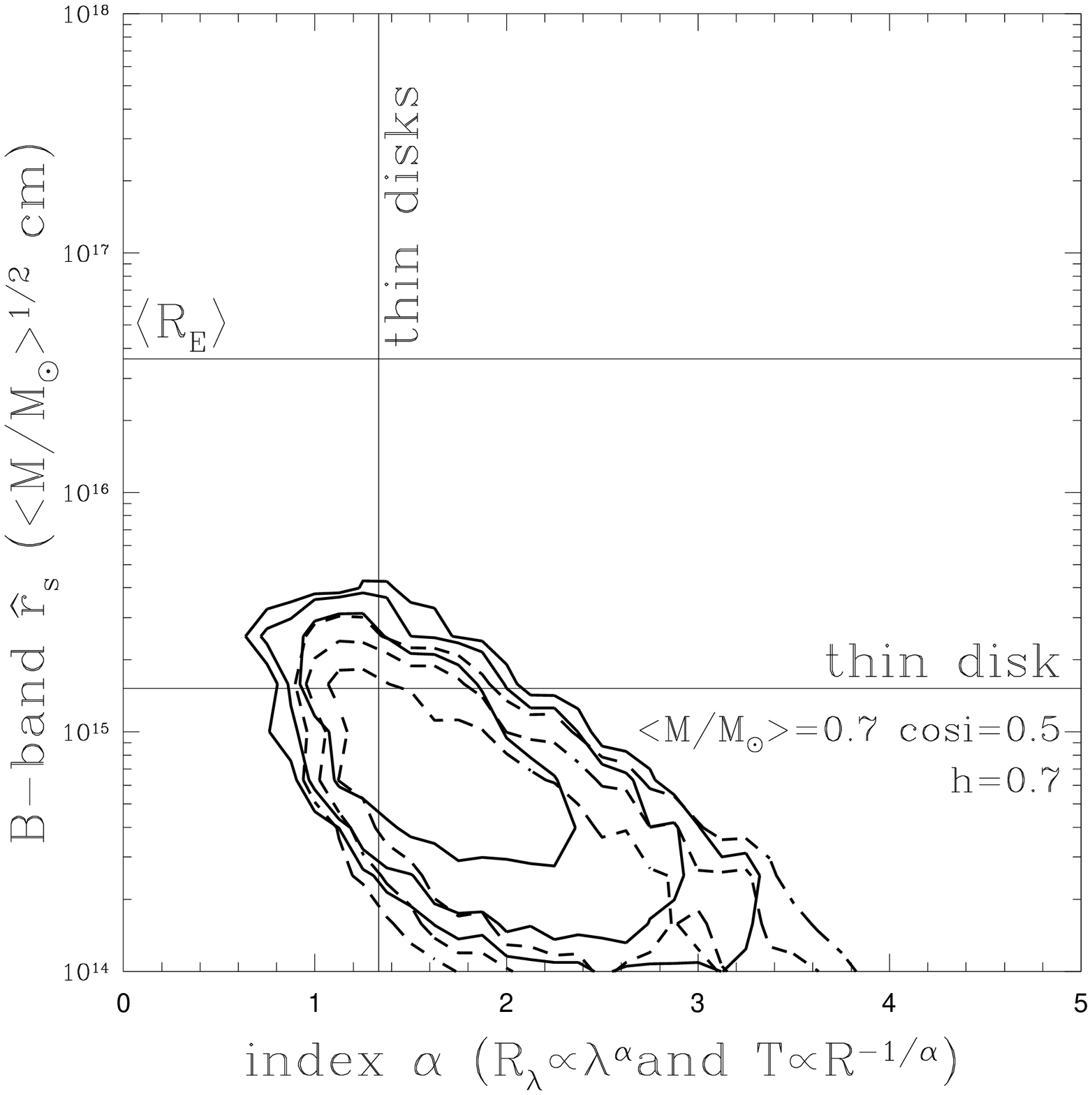,width=2.5in}
  \psfig{figure=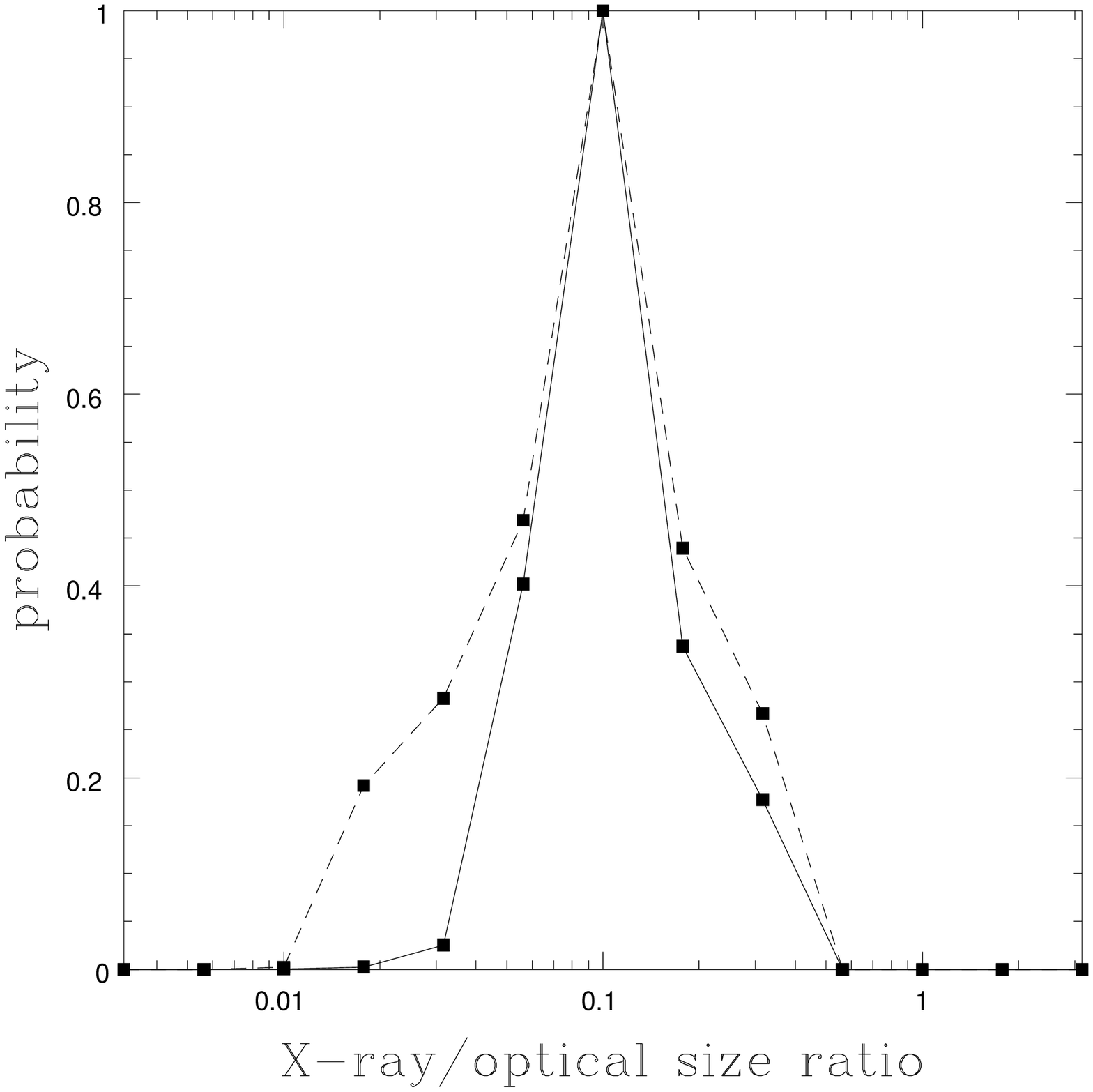,width=2.5in}
  }
\caption{ (Left) Likelihood contours for the scaling of the HE1104--1805 disk size with
   wavelength.  The contours are drawn at the 68\%, 90\% and 95\% enclosed likelihood
   contours for either a constant (solid) or logarithmic (dashed) prior on the disk
   size.  The lines indicate the mean Einstein radius $\langle R_E\rangle$, the 
   expected B-band disk size based on the Peng et al. (2006) estimate that 
   $\log (M_{BH}/M_\odot) = 9.4$, and the expected exponent $\alpha=4/3$ for
   the scaling of the size with wavelength, $R_\lambda \propto \lambda^{4/3}$.}
   \label{fig:he1104like}
\caption{ (Right) The probability distributions for the ratio of the X-ray and optical source sizes
  using either the first (dash) or both (solid) X-ray observations of RXJ1131.
  Note that the results are consistent, and that the error bars shrink with the addition of more data.  
  The formal uncertainties are approximately a factor of 1.7 ($\pm 0.23$ dex) using both epochs.
  These preliminary results assumed an SIE model for the lens which does not agree with the 
  measured time delays of the lens.
  }
\label{fig:rxj1131}
\end{figure}

\section{Preliminary Results}

We have just reached the point where our observations of microlensing in 
a large number of lenses and a over range of wavelengths have started to
produce results.  Here we present some examples of estimating the
stellar mass fraction, the mean stellar mass and the structure of
quasars as a function of black hole mass and wavelength.  
Let us start with an example of the data, in this case, a published 
V-band light curve of the Q2237+0305 image A/B flux ratio by the 
OGLE collaboration (Wozniak et al. 2000).  We used the OGLE data on this system in Kochanek (2004) to 
develop the analysis method and to demonstrate its power as a means of analyzing
microlensing data.  Fig.~\ref{fig:lcurve} shows the original light curve
we analyzed and one of the well-fitting model light curves 
produced by the Bayesian Monte Carlo analysis.  To illustrate the sensitivity
of light curves to source size, we also show the expected
light curves for sources of a different size from the V-band wavelength
used by OGLE.  

One interesting application of the Q2237+0305 microlensing data is shown
in Fig.~\ref{fig:kstar}, where we estimate the fraction $\kappa_*/\kappa$
of the surface density near the quasar images that is composed of stars.
This is one of the more difficult variables to measure from microlensing
data when the source size is also unknown, but the result is steadily 
converging to the expected result for Q2237+0305 of 
$\kappa_*/\kappa \simeq 1$.  We know the ``correct'' answer in this
case because the lens is the bulge of a 
low redshift spiral galaxy where we expect a negligible contribution to
the total surface density from dark matter.

We can also take a preliminary look at the mean mass of the microlenses
$\langle M \rangle$.  The mass is difficult to measure for any individual
lens because it depends on the square of the velocity, $\langle M \rangle \propto v_e^2$
and we have only
statistical information on the peculiar velocities of the lens.  As a 
result, we can probably never do better than a factor of $\sim 4$ 
uncertainty for $\langle M \rangle$ in any one lens.  With ensembles of lenses, however,
we can average out the uncertainties due to the peculiar velocities,
to produce an accurate estimate.  Fig.~\ref{fig:mass} shows a 
preliminary attempt at this, where we have averaged over the mass
estimates for eight lenses.  Formally, the median mass estimate is
$0.03 M_\odot$ with a 95.4\% confidence region of 
$ 0.01M_\odot < \langle M \rangle < 0.15 M_\odot$.  The result is
relatively robust.  For example if we drop the two lenses favoring the
lowest masses, the median rises to $0.08 M_\odot$ with 
$ 0.02M_\odot  < \langle M \rangle < 0.44 M_\odot$ at 95.4\%
confidence.  Such masses may be mildly inconsistent with local estimates of 
the mass function, as shown in Fig.~\ref{fig:mfunc}, where we compare
our Salpeter mass functions with these two normalizations to a 
detailed model for the mass function of the Galactic disk from
Gould (2000).  That the shapes differ is not important, as there
are extensive microlensing simulations demonstrating that only the
mean mass of the microlenses is easily measured (e.g. Paczy\'nski 1986,
Wyithe et al. 2000b), although we should probably shift to a shallower
mass function just to simplify comparisons.  The important point
is that our estimates of the mean mass are significantly lower than
the $\langle M \rangle =0.26M_\odot$ of the Galactic model.  On the
other hand, the low mass cutoff at $0.03 M_\odot$ in the Galactic model 
is somewhat arbitrary.

We now have enough well-sampled light curves that measurements of
disk sizes are becoming routine.  Fig.~\ref{fig:dsize}
shows a preliminary correlation of disk size with black
hole mass  $M_{BH}$.  The results track the expectations 
for thin disk theory from \S2 over two decades in black hole mass,
where the black hole mass estimates from Peng et al. (2006) are based
on the emission line widths of the quasars.  A formal fit to 
the estimates is marginally inconsistent with thin disk theory,
finding a slope of $R_\lambda \propto  M_{BH}^x$ 
that is somewhat steeper than the predicted $2/3$ (Eqn. 4)
and a higher normalization than expected for accretion at the Eddington
limit with an efficiency $\eta > 0.1$.  A potentially more serious
problem, also noted by Pooley et al. (2006), is that the disk sizes
are larger than the sizes predicted by matching a
black body emitting disk to the observed I-band magnitudes (Eqn. 6),
even though the microlensing and black body size estimates are 
proportional to each other.  The origin of this discrepancy is
probably due to a conspiracy of several smaller factors such
as contamination of the light curves by emission from larger
scale sources like the quasar broad lines, radiation transfer
problems in the disk atmosphere like line blanketing, 
scattering and reradiation of the disk emission, and 
differences in the advection of energy or the structure of 
real disks from the thin disk model.  Only the latter two
possibilities represent genuine changes from a thin disk.

One way to attack these problems of disk structure is to measure
disk sizes as a function of wavelength.
Fig.~\ref{fig:he1104lc} shows the wavelength-dependent
changes in the flux ratio of the two-image lens HE1104--1805,
and Fig.~\ref{fig:ximage} shows the dramatic variations in the
X-ray flux ratios of the lens RXJ~1131--1231.  Size ratios
can be measured more accurately than absolute sizes because
the ratios depend only on the relative variability amplitudes.  
Image A of HE1104--1805 has slowly switched 
from being bluer than image B to being redder in the optical/near-infrared 
while the mid-infrared fluxes agree with the flux ratio of the broad 
emission lines (e.g. Wisotzki et al. 1995).  An unexpected feature of the wavelength
ordering, although one which we had no difficulty reproducing,
is the present situation where A is brightest relative
to B in the J band and becomes relatively fainter at both bluer
and redder wavelengths.
If we fit these light curves with a model in which the size of 
the disk is a power-law referenced to the size in the B-band,
$R_\lambda = R_B (\lambda/\lambda_B)^\alpha$, we find the 
likelihood distributions shown in Fig.~\ref{fig:he1104like}.
The results are consistent with the expectations of thin disk
theory given the estimated black hole mass from Peng et al. (2006),
but the exact results depend on the prior adopted for the 
source size.  The incomplete convergence is probably due to
the very limited data in the redder bands in the early phases
of the light curves.

We can also estimate the size ratio between the X-ray emitting regions
and the optical disk by simultaneously fitting the optical light curve
and sparse X-ray measurements.  Fig.~\ref{fig:rxj1131} shows
that the X-ray source in RXJ~1131--1231 is roughly an order
of magnitude more compact than the optical source, with
remarkably small uncertainties given that the estimate is based
on only two X-ray observations.  If X-ray emission comes from
near the last stable orbit of the black hole, as our estimates
so far have suggested, then we should see a large optical/X-ray
size ratio in a system like RXJ1131--1231 with a relatively low
mass black hole ($M_{BH} \simeq 10^{7.8}M_\odot$) and a smaller 
size ratio in a system like Q2237+030 with a relatively high
mass black hole ($M_{BH}\simeq 10^{9.0}M_\odot$) 
because the disks of lower mass black holes are hotter and
the observed-frame optical emission comes from farther out
in the disk (see Fig.~\ref{fig:dsize}).  A preliminary analysis of our X-ray observations
of Q2237+030 agrees with this expectation.

\section{Physical, Computational and Statistical Issues}

In this section we discuss aspects of the physical models, the computational 
methods and the statistical analysis that need to be considered.  
Generally, they are not independent -- physical issues
may arise because of computational limitations and so forth.

\subsection{Physical Issues}

We are principally worried about four physical issues in the way we
analyze the problem, which we discuss in order from very mildly
worried to more seriously worried.  We have experimented with most
of the issues, and for the moment they appear not to be a limiting
problem compared to the difficulty in obtaining the data, but this
will change as the data improves.

First, in our standard models we assume a Salpeter mass function with a dynamic
range of a factor of 100 in the mass.  We have experimented with
smaller mass ranges or even single masses, and we have been 
unable to distinguish between mass functions or find significant
differences in the results for other variables from changes in the
mass functions.  This experience matches that from extensive 
theoretical studies of the effect of the mass function over the
last 20 years (e.g. Wyithe et al. 2000b).  Reasonable mass functions have most of the mass
in a limited range, and the physical scales depend only on the
square root of the mass, so this insensitivity is not surprising particularly given
the significant uncertainties in results based on the available data.
Eventually, however, it may matter.  

Second, we can choose to fit only the time variations in the image flux ratios
that are the easily observed signature of microlensing, or to also fit
the actual flux ratios.  If you examine the magnification patterns in
Fig.~\ref{fig:map}, you see that the pattern is typically divided into
broad, relatively smooth regions of demagnification separated by 
narrow regions of higher magnification consisting of dense networks
of caustics.  A given amplitude of variability can be
created by moving slowly through a region of high magnification with
a large source or by moving rapidly through a region of low magnification with
a smaller source.  Any information that constrains you to a 
particular region of the maps can greatly reduce degeneracies in the
solutions.  

Forcing the models to match the flux ratios predicted by the ``macro'' model for
the overall mass distribution is such a constraint.  The ``macro''
model is already part of the microlensing model because it is the
source of the estimated values of the surface density $\kappa$ and 
shear $\gamma$ near each image used in generating the patterns.  
There are two problems with forcing the models to match the observed flux ratios.
One problem is that undetected satellite galaxies (``substructures'')
near the lensed images can change the image magnifications while
having little other detectable effect, so it is risky to force
the flux ratios to match that of a standard ``macro'' model (e.g. 
Kochanek \& Dalal 2004).  These
satellites may also mean that the estimates of $\kappa$ and $\gamma$
used to generate the magnification maps are sometimes incorrect.
A second problem is that the observed flux ratios, but probably 
not their time variations, can be affected by absorption in the 
interstellar medium of the lens galaxy (e.g. Falco et al. 1999).  
In short, adding the absolute flux
ratios as a constraint would be very useful, but requires a clear 
understanding of these other systematic problems.

Third, our current disk model is a face-on thin disk without the
central temperature depression created by the inner edge of the accretion
disk (Eqns.~\ref{eqn:tdisk}-\ref{eqn:fnu}).  At fixed wavelength, neglecting 
the inner edge has considerably less consequence than one 
might naively expect -- at fixed wavelength little flux is radiated there 
and the finite resolution of the magnification patterns eliminates the formal
surface brightness divergence in Eqn. (5). 
We have experimented with adding the hole or simply using a Gaussian
and we cannot statistically distinguish the models with the present data.  This
is again the expectation from earlier theoretical work -- microlensing
is primarily sensitive to an effective smoothing area, and the resulting
size estimate is only weekly sensitive to the true surface brightness profile (e.g.
Mortonson et al. 2005).
Once we can distinguish profiles, our problems begin, because the space
of models increases dramatically once we move away from this simple model.
Even within the context of the standard thin disk, there is the scale of the
inner disk edge ($R_{in}$) and the disk inclination.  For more complex
disk models, there are relativistic corrections to the 
observed temperatures and photon trajectories,  
effects from variable scale heights coupled with the viewing angle, 
more realistic radiation transfer out of the disk, non-thermal emission 
mechanisms, and the inner disk may be a thick torus rather than
a standard thin disk (e.g. Li et al. 2005). 

While distinguishing different spatial structures for the disk simply by fitting
the light curve measured in a single band, determining the structure of the disk 
as a function of wavelength is relatively straight forward.  Smaller sources have larger
variability amplitudes than bigger sources, so the signature of size changes
is unambiguous.  Additionally, ratios of source sizes are dimensionless 
quantities that are unaffected by uncertainties in the mass scale 
$\langle M \rangle$, so we can determine size ratios more
accurately than the absolute size scale.  The changes in size over the
optical/near-IR wavelength baseline are quite large --   
if $R_\lambda \propto \lambda^{4/3}$ (Eqn. 4), the disk size changes
by a factor of $3.4$ over our J-band to B-band wavelength baseline,
which corresponds to over an order of magnitude change in the area of
the source. 

Finally, the issue we are most concerned about is the use of a static magnification
pattern where the stars do not move.  The concern is not that
we are getting incorrect results, but that the compromises from
using static patterns force us to results with broader uncertainties
than we would have if we allowed dynamic patterns.   
We are forced to this approximation by computational issues -- as we
discuss in \S6.2.

If you examine the maps in Fig.~\ref{fig:map}, you can see that
the maps have different scales in the strongly and weakly sheared
directions, so the velocity required to produce a given 
amount of variability depends on the direction of the 
motion.  The  source velocity vector is the
same for all images, so we should see different statistics for the
variability of the images depending on the angle between the source
velocity and the strongly sheared direction of the magnification 
patterns.  If we use a fixed pattern and have no random stellar motions, 
it is unreasonable to force a fixed source velocity -- instead, we use 
a fixed velocity magnitude 
but allow the direction on each pattern to be random.
Another problem with fixed patterns is that we lose the contribution
of ``pattern velocities'' created by rearrangements of the caustics 
as the stars move.  Pattern velocities can be very
much higher than physical velocities, and may be the 
source of the rapid microlensing variability observed in HE1104--1805
(Schechter et al. 2003).

\subsection{Computational Issues}

There are three general computational issues: (1) building magnification
maps, (2) fitting them into computer memory, and (3) rapidly generating the
results for trial light curves.

We generate the magnification maps by ``shooting'' a uniform grid of
light rays through a star field, bending the trajectory based on the
gravitational fields of the stars and then counting where the rays
land on the source plane (Kayser et al. 1986, Schneider \& Weiss 1988). 
The magnification at point on the source plane is the ratio between the 
input ray density and the density of rays reaching that point on the 
source plane.  The problem is that stars,
as point masses, have divergent ray deflections at small impact 
parameters, so stars very distant in projection from a point in the 
magnification map may contribute to the result.  Traditionally the
solution has been to populate a large region with stars in order
to determine the source magnification maps for a much smaller 
region.  This is relatively inefficient, and it leads to magnification
maps with edges that make it difficult to generate random light curves
because you would have to avoid any edge-crossing path.
We solve this problem by using source and lens plane regions both of
which are periodic and correspond to projections of each other.  
In essence, we shoot the rays through an infinite, if periodic, star field, and
the resulting magnification patterns have no
edges.  In order to make both the lens and source planes periodic,
we must adjust the desired shear $\gamma$ by an amount of order
the inverse of the array dimensions ($<10^{-3}$), but this is
a small enough to have no physical consequences. 

The next problem is memory. The magnification maps must have a
large enough outer scale to be statistically representative of
the star field and to allow the production of large numbers of
independent light curves.  They must also have a small enough
pixel scale  to resolve the gravitational radius of the quasar 
black hole.    Typically a $4096^2$ grid scale is adequate to
cover a region with an outer scale of $20$-$40 R_E$ and an inner
scale that resolves the disk.    A single $4096^2$ magnification 
pattern is only 67~MByte, or 268~Mbyte
for four images.  For a single light curve, this is all that is
necessary because models for different source sizes can be run
independently.  For models with multiple light curves, each
requiring a different source size, the convolved patterns
for all source sizes must be in memory --
a four-image lens with 10 source sizes (2.7~GByte)
will fit on a standard computer, but
it is beginning to push the limits. The big problems arise when
we allow the stars to move and we need an animated sequence of
magnification patterns.  This can easily require 100 independent
magnification patterns for each image.  Add 
multiple source sizes, and the memory requirements quickly
approach several hundred Gbytes.  The present version of the
code is multi-threaded, so such large problems can be run efficiently
on shared-memory supercomputers, but it is not parallelized, so
it cannot be run on clusters.  Parallelization may not
be effective because we must do random look-up
of points in magnification maps spread across the processors.

The remaining computational issue is simply that of speed when faced
with performing the necessary Bayesian integrals over all the 
nuisance and physically interesting variables.  We use several
tricks to rapidly find regions producing good fits to the data,
and these raise some of the statistical issues discussed in the
next section.  Our first trick is to only use trial light curves
that fit the data better than a threshold $\chi^2$, usually a few 
times the number of degrees of freedom $N_{dof}$, in the Bayesian 
integrals.  The advantage of doing this is that rather than fitting 
the whole light curve we fit it as a sequence of randomly ordered 
points and then discard the trial if the goodness of fit ever exceeds the 
threshold.  This leads to an enormous gain in execution speed 
because many trials are thrown away after fitting a very small
number of points.  To minimize biases, we periodically generate new 
random orderings of the light curve.  The second trick is that for light curves
that pass the threshold we optimize the initial position of the source, 
the direction of motion and allow for small fractional changes in the 
velocity.  Only small changes in the velocity are allowed because we want 
the velocity to change by significantly less than the scales on which we will
eventually bin the velocities for later analysis. Essentially,
we have a hidden, localized maximum likelihood loop for the 
nuisance variables.  

\subsection{Statistical Issues}

As we have currently implemented the code, there are several lingering
statistical issues.

We start with a $\chi^2$ statistic as our estimate for the probability that
a particular trial light curve $i$ fit the data, $P(D|p) = \exp(-\chi_i^2/2)$.  When we have
a light curve with significant numbers of data points, the resulting
probabilities are very sensitive to whether we have correctly estimated
the uncertainties, systematic as well as statistical, in the light curves.
Small changes in the estimated uncertainties dramatically reweight the
probabilities.  We deal with this uncertainty by assuming that the
photometric errors themselves have some uncertainties, characterized
by another nuisance parameter, and then averaging the probabilities
over a range for this variable.  

For the computational reasons outlined above, we have used fixed 
magnification patterns in our current analyses rather than allowing
the stars to move.  We compensate for this simplification in two
ways.  First, we allow the direction of motion 
to be a random independent variable {\it for
each image.}  This is clearly incorrect for the bulk velocities,
but it should fail in a statistically conservative manner by 
producing broader probability distributions.  It should, 
however, partly mimic the effects of random motions of the 
stars.  Second, in the prior for the effective velocity we
must somehow mimic the effect of the random velocities of 
the stars.  A random bulk velocity, like the peculiar velocity
of the lens galaxy, has a non-trivial probability of values
much smaller than the rms peculiar velocity.  Similarly,
each star in the pattern has a non-trivial probability of having
a velocity that is small compared to the velocity dispersion
of the lens.  The star field creating the magnification
pattern includes the effects of many stars and so the
``statistical'' velocities of the stars creating the 
pattern are unlikely to differ significantly from the velocity
dispersion of the stars.  We can mimic this physical effect
statistically, but it will introduce some biases (Wyithe et al. 2000a)
that should be small compared to our present uncertainties.  


The trick introduced to speed up the calculations means that
we have cheated slightly in doing the Bayesian Monte Carlo integrals.  
When we drop a trial because it does not reach the $\chi^2$
threshold, it is lost forever because we never 
calculate its true goodness of fit.  While the probability
associated with such light curves $P(D|p)=\exp(-\chi^2/2)$ is negligible
compared to one producing a good fit, they make a contribution
to the Bayesian integrals of $n(\chi^2) \exp(-\chi^2/2)$ which is 
the low probability multiplied by the number of such trials $n(\chi^2)$.
We need to be sure that this total contribution of poorly fitting
trials still has no statistical weight in the Bayesian integrals.  
Experiments with adjusting the threshold indicates that
this is not an important issue. 

The inner optimization loop is the last issue.  Essentially,
we take a small volume in a sub-space of the parameters and
adopt the localized maximum likelihood solution rather than 
the full Bayesian integral over that sub-space.  A better
way of performing this step would be to also calculate a 
statistical weight for the volume sampled by the optimization, 
or to approximate this from the curvature of the $\chi^2$ near 
the maximum (i.e. a local Laplace approximation). That being said, 
we have not seen any difference in the results from using
this procedure versus dropping the local optimization.


\section{Summary}

We have reached the point where we can begin to use microlensing
as an industrial tool.  The data allows us to study problems -- stellar mass 
fractions, mean stellar masses and quasar structure -- that
are virtually impossible to probe with other methods.  
Our main bottleneck now is time.  Time matters on two levels.  The
first issue is simply observing time.  We have enough to 
monitor approximately 20 lenses well at one wavelength (R-band),
and to obtain sparse coverage at other wavelengths (J,I,V and B-bands) 
and occasional data from HST and Chandra at X-ray and UV wavelengths 
blocked by the atmosphere.  The second issue is that the characteristic 
time scales for microlensing are long -- roughly a decade for any 
particular image.  For good statistical results we need to accumulate
``image monitoring centuries'' either by monitoring many lenses for
shorter periods of time or a few lenses for very long periods of 
time.  Given that all the results to date are being obtained with
roughly 10\% of the time on a 1.3m telescope, there is enormous 
room to accelerate the scientific program.  All it requires is
a few more queue-scheduled telescopes, hopefully in the 2-meter
class.   

Some of the most interesting problems require greater use of 
HST and Chandra because we expect the UV and X-ray emission 
of the accretion disk to come from the regions closest to the
black hole.  In particular, the microlensing of wavelengths 
shortward of the 3500\AA\ atmospheric cutoff should show
significant differences between the lower mass systems
($M_{BH} \sim 10^8M_\odot$) and the high mass systems
($M_{BH} \sim 10^{9}M_\odot$) because they correspond to 
regions still well outside the inner edge of the disk for
the low mass systems but to regions very close to the
inner edge for the high mass systems (see Fig.~\ref{fig:dsize}).  The X-ray observations
need to be extended from short observations sufficient to
measure broad band variability, to long observations where
we can examine the microlensing of the X-ray spectrum, particularly
the Iron K$\alpha$ lines.  Existing data show evidence for
microlensing of the lines relative to the continuum
(e.g. Dai et al. 2003), but better data is required to 
produce clean results.

\acknowledgements 
The authors would like to thank G. Garmire for using some of his guaranteed CXO time
to make some of these observations, and M. Dietrich, P. Osmer, B. Peterson, and R. Pogge
for extensive discussions about quasar structure.
This work is based on observations made with the Spitzer Space Telescope, which is operated 
by the Jet Propulsion Laboratory, California Institute of Technology under a contract with NASA. 
Support for this work was provided by NASA through Cycle 2 award 20451 issued by JPL/Caltech.
Support for this work was provided by the National Aeronautics and Space Administration through Chandra 
Award Number 6836 issued by the Chandra X-ray Observatory Center, which is operated by the 
Smithsonian Astrophysical Observatory for and on behalf of the National Aeronautics Space Administration 
under contract NAS8-03060."


\end{document}